\documentclass[a4paper, amsfonts, amssymb, amsmath, reprint, showkeys, footinbib, twoside,superscriptaddress,floatfix,longbibliography]{revtex4-1}
\usepackage{tabularx}
\usepackage{booktabs}
\usepackage{placeins}
\usepackage{multirow}
\usepackage{graphicx}%
\usepackage{dcolumn}%
\usepackage{bm}%
\usepackage{float}
\usepackage{xcolor}
\usepackage{mhchem}
\usepackage{gensymb}
\usepackage[slantedGreek]{newtxmath}
\usepackage{comment}
\usepackage[pdftex, bookmarks, pdffitwindow=false, pdfstartview=FitH, pdfdisplaydoctitle, colorlinks, plainpages=false, pdftitle={},pdfauthor={}, pdfpagelabels, hypertexnames, citecolor={blue!50!black},linkcolor={blue!50!black}, urlcolor={blue!50!black}, pdflang={en}, hyperfootnotes=false, breaklinks]{hyperref}
\usepackage{siunitx}
\usepackage{subcaption}

\newcommand{\kb}{k_{\rm B}} %

\newcommand{\figLabelCapt}[1]{\textbf{\MakeLowercase{{#1}}}}
\newcommand{\refSub}[2]{\hyperref[#2]{\ref{#2}\figLabelCapt{#1}}}
\newcommand{\figref}[1]{Fig.~\ref{#1}}
\newcommand{\figrefsub}[2]{Fig.~\refSub{#2}{#1}}
\newcommand{\figsrefsub}[2]{Figs.~\refSub{#2}{#1}}

\usepackage[justification=raggedright,singlelinecheck=false]{caption}

\makeatletter
\def\@bibdataout@aps{
 \immediate\write\@bibdataout{
 @CONTROL{
   apsrev41Control, author="48",editor="1",pages="0",title="0",year="1"
 }}
 \if@filesw
  \immediate\write\@auxout{\string\citation{apsrev41Control}}
 \fi
}
\makeatother

\begin{document}

\title{Chemical potentials from structure factors: II. Charged multi-component mixtures}

\author{Xiaoyu Wang}
\thanks{These authors contributed equally.}
\affiliation{Department of Chemistry, UC Berkeley, California, 94720, United States}

\author{Roya Savoj}
\thanks{These authors contributed equally.}
\affiliation{Department of Chemistry, UC Berkeley, California, 94720, United States}

\author{Musahid Ahmed}
\affiliation{Chemical Sciences Division, Lawrence Berkeley National Laboratory, Berkeley, California, 94720, United States}

\author{Bingqing Cheng}
\email{bingqingcheng@berkeley.edu}
\affiliation{Department of Chemistry, UC Berkeley, California, 94720, United States}
\affiliation{Chemical Sciences Division, Lawrence Berkeley National Laboratory, Berkeley, California, 94720, United States}
\affiliation{Bakar Institute of Digital Materials for the Planet, UC Berkeley, California, 94720, United States}

\date{\today}

\begin{abstract}
The chemical potentials of charged multi-component mixtures are central to electrolyte thermodynamics, but remain difficult to compute from atomistic simulations. The S0 method enables computing chemical potentials of mixtures from equilibrium molecular dynamics simulations. Here we extend the S0 method to charged mixtures by combining the composition-space framework developed in Part I: Neutral Multi-component Mixtures with a Coulombic treatment of the small-wavenumber limits of static structure factors. This approach separates thermodynamically relevant neutral composition fluctuations from forbidden macroscopic charge fluctuations, and further accounts for charge-neutrality constraints. We use the method to compute the chemical potentials of multiple-halide aqueous salt solutions, elucidating the ion-specific thermodynamic effects. We also calculate the mixing free energies of molten salt mixtures, demonstrating the importance of correctly describing long-wavelength electrostatic correlations.
\end{abstract}

\maketitle

\section{Introduction}

The chemical potentials of charged mixtures underlie many thermodynamic phenomena, including salt activities, solvation, ion exchange and transport~\cite{robinson2002electrolyte,Pyeongeun2024Interfacial}. 
Computing chemical potentials in atomistic simulations remains challenging even for simple electrolyte solutions~\cite{nezbeda2016recent}. Conventional free-energy perturbation and thermodynamic integration methods~\cite{Li2017,Li2018,defever2021computing,shah2023first} typically require sampling multiple intermediate states and can suffer from slow convergence~\cite{allen2012computer,gibson2025computing}. When such intermediate states have net charges, they demand additional treatment of periodic electrostatics and are more sensitive to finite-size effects~\cite{Saravi2021Individual}. 
Monte Carlo approaches can preserve charge neutrality through semigrand-canonical identity exchanges~\cite{xie2023semigrand} or osmotic-ensemble insertion of neutral ionic groups~\cite{lisal2005molecular,moucka2011molecular}, but these methods often suffer from low acceptance ratios and convergence issues in dense ionic systems~\cite{perego2016chemical,moucka2015electrolyte}. 
Chemical potential derivatives can alternatively be obtained from molecular dynamics (MD) simulations using the Kirkwood-Buff integrals (KBIs)~\cite{Kirkwood1951}, but direct evaluation of KBIs in real space requires appropriate finite-system corrections as addressed in recent methodological developments~\cite{CortesHuerto2016,kruger2018size,Dawass2018KB,Dawass2019KB,Chattopadhyay2025KB,robustBusselez2025,Peroutka2026molecular}. 
To avoid direct KBI evaluations,
the S0 method computes the chemical potential derivatives from the small-wavenumber limits of partial static structure factors~\cite{cheng2022computing}.

In Part I: Neutral Multi-component Mixtures~\cite{savoj2026computing}, we extended the original two-component S0 method~\cite{cheng2022computing} to multi-component mixtures. 
For neutral mixtures, the partial structure factors $S(\mathbf k)$ are extrapolated to the long-wavelength ($k\rightarrow 0$) limit using the Ornstein--Zernike form~\cite{barrat2003basic,hansen2013theory}, and these $S^0$ values can be used to construct the derivative matrix that is subsequently integrated over composition using a Gaussian process (GP) to obtain chemical potentials of individual components. 

However, the neutral S0 formulation is not directly applicable to charged systems.
Importantly, the long-range Coulomb interaction changes the structure-factor extrapolation:
Macroscopic charge fluctuations are suppressed, and the charge-density structure factor follows the Stillinger--Lovett limiting behavior~\cite{lee1997charge,stillinger1968ion}, $S_{ZZ}(\mathbf{k})\sim k^2$, approaching zero at small $k$ values. 
The ordinary Ornstein--Zernike form~\cite{barrat2003basic,hansen2013theory} therefore no longer applies.
In addition, the accessible composition space is restricted not only by the normalization of particle number fractions, $\sum_i x_i = 1$, but also by charge neutrality. 
Finally, individual ionic chemical potentials are not uniquely defined without specifying an electrostatic convention; experimentally meaningful thermodynamic quantities are neutral combinations of ionic chemical potentials, such as salt chemical potentials.

Here, we present Part II. Charged multi-component mixtures, which develops a Coulomb-aware small-$k$ structure-factor representation and formulates a projection to compute chemical potential derivatives under charge neutrality.
The framework
separates thermodynamically relevant neutral composition fluctuations from the suppressed
macroscopic charge mode. We then apply the method to multiple-halide aqueous electrolyte solutions and to
mixtures of molten salts.

\section{Theory}

\subsection{Long-wavelength density fluctuations}

To derive the functional form of structure factors at small $k$ for general cases, 
we start with a weakly-inhomogeneous single-component fluid with $N$ particles in volume $V$. 
Let $\rho(\mathbf{r})=\rho_0+\Delta\rho(\mathbf{r})$ be a slowly varying density field around a uniform reference state. 
The Fourier component of the density fluctuation is
\begin{equation}
    \Delta\rho(\mathbf{k}) =
    \int_V d\mathbf{r}\,\Delta\rho(\mathbf{r})\exp(-i\mathbf{k}\cdot\mathbf{r}),
\end{equation}
and the static structure factor is
\begin{equation}
    S(\mathbf{k}) =
    \frac{1}{N}
    \left<
    \Delta\rho(\mathbf{k})\Delta\rho(-\mathbf{k})
    \right>.
\end{equation}
For small-amplitude, long-wavelength fluctuations, linear response gives the quadratic free energy
\begin{equation}
    F[\rho_0+\Delta\rho]
    =
    V f_0(\rho_0)
    +
    \frac{\kb T}{2\rho_0 V}
    \sum_{\mathbf{k}}
    \frac{
    \Delta\rho(\mathbf{k})\Delta\rho(-\mathbf{k})
    }{
    S(\mathbf{k})
    }
    +
    \mathcal{O}(\Delta\rho^3).
    \label{eq:single_response}
\end{equation}
In a neutral fluid with only short-range interactions, the same free energy functional may be expanded in a square-gradient form,
\begin{equation}
    F_{\mathrm{SR}}[\rho]
    =
    \int d\mathbf{r}
    \left[
    f_0(\rho(\mathbf{r}))
    +
    \frac{\kb T\,\xi_0^2}{2\rho_0}
    |\nabla\rho(\mathbf{r})|^2
    +\cdots
    \right].
    \label{eq:F_SR_single}
\end{equation}
Comparison with Eqn.~\eqref{eq:single_response} yields the usual Ornstein--Zernike (OZ) small-$k$ form~\cite{barrat2003basic,hansen2013theory}
\begin{equation}
    S^{\mathrm{OZ}}(\mathbf{k})
    =
    \frac{S^{\mathrm{OZ}}(0)}{1+k^2\xi^2},
    \label{eq:single_oz}
\end{equation}
where $\xi$ is the corresponding correlation length,
$\xi^2=\kb T (\rho_0\,f_0'')^{-1} \xi_0^2$.

\subsection{Bare-Coulomb correction}

If the single-component fluid is made of charged particles of valence $z$ and immersed in a uniform neutralizing background, the short-range functional must be supplemented by the electrostatic energy associated with charge-density fluctuations:
\begin{equation}
    F[\rho] =
    F_\mathrm{SR}[\rho]
    + F_\mathrm{LR}[\rho],
\end{equation}
\begin{equation}
F_\mathrm{LR}[\rho] =
\frac{z^2}{2}
\iint d\mathbf r\, d\mathbf r'\,
v_\mathrm{LR}(\mathbf r,\mathbf r')\,
\Delta \rho(\mathbf r)\,
\Delta \rho(\mathbf r').
\label{eq:F_w_LR}
\end{equation}
The interaction kernel $v_\mathrm{LR}(\mathbf r,\mathbf r') = 1/ (4 \pi \varepsilon_0 |\mathbf r-\mathbf r'|) $ in Eqn.~\eqref{eq:F_w_LR} is the bare-Coulomb (BC) kernel, not a screened Debye--Huckel Green's function,
because the mobile ions remain explicit fluctuation variables in the structure factor.
In Fourier space, 
\begin{equation}
    v_{\mathrm{LR}}(\mathbf{k})
    =
    \frac{1}{\varepsilon_0 k^2},
    \label{eq:bare_coulomb_kernel}
\end{equation}
and the $k=0$ mode is excluded by overall charge neutrality.

Combining the short-range square-gradient contribution in Eqn.~\eqref{eq:F_SR_single} with Eqn.~\eqref{eq:bare_coulomb_kernel} gives
\begin{equation}
    \frac{\kb T}{\rho_0}
    \frac{1}{S^{\mathrm{BC}}(\mathbf{k})}
    =
    f_0''(\rho_0)
    +
    \frac{\kb T k^2\xi_0^2}{\rho_0}
    +
    \frac{z^2}{\varepsilon_0 k^2}.
\end{equation}
Introducing the Debye wavevector $\kappa_D$, which is the inverse of the Debye length, $\kappa_D = 1/\lambda_D$,
\begin{equation}
    \kappa_D^2
    =
    \frac{\rho_0 z^2}{\varepsilon_0\kb T},
\label{eq:kappa_pure}
\end{equation}
one obtains
\begin{equation}
    S^{\mathrm{BC}}(\mathbf{k})
    =
    \frac{k^2}{
    \kappa_D^2
    +
    a k^2
    +
    b k^4
    },
    \label{eq:single_bc}
\end{equation}
where $a$ and $b$ collect the short-range compressibility and square-gradient terms. 
Thus a pure charge-density fluctuation has $S^{\mathrm{BC}}(\mathbf{k})\rightarrow0$ as $k\rightarrow0$, consistent with macroscopic charge neutrality and the Stillinger--Lovett second-moment condition~\cite{lee1997charge,stillinger1968ion}. Physically, this means that increasingly long-wavelength fluctuations cannot carry a net charge.

\subsection{Multi-component matrix form}

We now generalize this construction to a mixture with multiple components indexed by $i$. 
Let $\rho_0=\sum_i\rho_i$ be the total number density, and $x_i=\rho_i/\rho_0$ the composition.
We define
\begin{align}
    \Delta\boldsymbol{\rho}(\mathbf{k})
    &=
    \left(
    \Delta\rho_1(\mathbf{k}),
    \Delta\rho_2(\mathbf{k}),\ldots
    \right)^T,
    \\
    \mathbf{Y}
    &=
    \operatorname{diag}(\sqrt{x_1},\sqrt{x_2},\ldots).
\end{align}
Like Eqn.~\eqref{eq:single_response}, the matrix of partial structure factors is defined through the quadratic response
\begin{equation}
    F
    =
    V f_0
    +
    \frac{\kb T}{2V\rho_0}
    \sum_{\mathbf{k}}
    \Delta\boldsymbol{\rho}^{T}(-\mathbf{k})
    \left[
    \mathbf{Y}^{-1}
    \mathbf{S}^{-1}(\mathbf{k})
    \mathbf{Y}^{-1}
    \right]
    \Delta\boldsymbol{\rho}(\mathbf{k}).
    \label{eq:matrix_response}
\end{equation}
Extending Eqn.~\eqref{eq:F_SR_single}, the short-range free energy functional becomes
\begin{equation}
    F_{\mathrm{SR}}
    =
    V f_0
    +
    \frac{\kb T}{2V\rho_0}
    \sum_{\mathbf{k}}
    \Delta\boldsymbol{\rho}^{T}(-\mathbf{k})
    \left[
    \mathbf{H}
    +
    k^2\mathbf{Y}^{-1}\mathbf{L}\mathbf{Y}^{-1}
    \right]
    \Delta\boldsymbol{\rho}(\mathbf{k}),
    \label{eq:matrix_sr}
\end{equation}
where
\begin{equation}
    H_{ij}
    =
    \frac{\rho_0}{\kb T}
    \frac{\partial^2 f_0}{\partial\rho_i\partial\rho_j},
\end{equation}
and $\mathbf{L}$ is a matrix of square-gradient coefficients.
Comparison of Eqns.~\eqref{eq:matrix_response} and~\eqref{eq:matrix_sr} gives the neutral-mixture Ornstein--Zernike matrix form
\begin{equation}
\boxed{
    \mathbf{S}^{\mathrm{OZ}}(\mathbf{k})
    =
    \left[
    \mathbf{Y}\mathbf{H}\mathbf{Y}
    +
    k^2\mathbf{L}
    \right]^{-1}
    }.
    \label{eq:matrix_oz}
\end{equation}
This is the form used in neutral mixtures to obtain the $k\rightarrow0$ structure-factor limit in Part I~\cite{savoj2026computing}.

For a charged mixture with valences $\mathbf{z}=(z_1,z_2,\ldots)^T$, the long-range Coulomb contribution is 
\begin{equation}
    F_{\mathrm{LR}}
    =
    \frac{1}{2V}
    \sum_{\mathbf{k}\neq0}
    \Delta\boldsymbol{\rho}^{T}(-\mathbf{k})
    \frac{
    \mathbf{z}\mathbf{z}^{T}
    }{
    \varepsilon_0 k^2
    }
    \Delta\boldsymbol{\rho}(\mathbf{k}).
    \label{eq:matrix_flr}
\end{equation}
With the multi-component Debye wavevector,
\begin{equation}
    \kappa_D^2
    =
    \frac{\rho_0}{\varepsilon_0\kb T}
    \mathbf{z}^{T}\mathbf{Y}^2\mathbf{z},
    \label{eq:matrix_kappa}
\end{equation}
and combining Eqns.~\eqref{eq:matrix_response}, \eqref{eq:matrix_sr}, and~\eqref{eq:matrix_flr}, the Coulomb-aware partial structure-factor matrix is
\begin{equation}
\boxed{
    \mathbf{S}^{\mathrm{BC}}(\mathbf{k})
    =
    \left[
    \mathbf{Y}\mathbf{H}\mathbf{Y}
    +
    k^2\mathbf{L}
    +
    \frac{
    \mathbf{Y}\mathbf{z}\mathbf{z}^{T}\mathbf{Y}
    }{
    \mathbf{z}^{T}\mathbf{Y}^2\mathbf{z}
    }
    \frac{\kappa_D^2}{k^2}
    \right]^{-1}
    }.
    \label{eq:matrix_bc}
\end{equation}
Using the Sherman--Morrison identity~\cite{sherman1950adjustment}, the finite limiting matrix is
\begin{equation}
\boxed{
    \mathbf{S}^{\mathrm{BC}}(0^+)
    =
    \mathbf{A}^{-1}
    -
    \frac{
    \mathbf{A}^{-1}\mathbf{w}\mathbf{w}^{T}\mathbf{A}^{-1}
    }{
    \mathbf{w}^{T}\mathbf{A}^{-1}\mathbf{w}
    }},
    \label{eq:matrix_bc_zero}
\end{equation}
where
\begin{equation}
    \mathbf{A}
    =
    \mathbf{Y}\mathbf{H}\mathbf{Y},
    \qquad
    \mathbf{w}
    =
    \mathbf{Y}\mathbf{z}.
\end{equation}
Eqns.~\eqref{eq:matrix_bc} and ~\eqref{eq:matrix_bc_zero} are then used together for determining the $k\rightarrow0$ structure-factor limit for charged mixtures.
$\kappa_D$ can either be determined by the physical charged particle densities or be treated as a fitting parameter.

Eqn.~\eqref{eq:matrix_bc_zero} shows explicitly that the charge mode is projected out, as the corresponding normalized charge-density structure factor is
\begin{equation}
    S_{ZZ}^{\mathrm{BC}}(\mathbf{k})
    =
    \frac{
    \mathbf{z}^{T}\mathbf{Y}
    \mathbf{S}^{\mathrm{BC}}(\mathbf{k})
    \mathbf{Y}\mathbf{z}
    }{
    \mathbf{z}^{T}\mathbf{Y}^2\mathbf{z}
    }.
    \label{eq:szz_def}
\end{equation}
With 
\begin{equation}
    m(\mathbf{k})
    =
    \frac{
    \mathbf{w}^{T}
    \left[
    \mathbf{Y}\mathbf{H}\mathbf{Y}
    +
    k^2\mathbf{L}
    \right]^{-1}
    \mathbf{w}
    }{
    \mathbf{w}^{T}\mathbf{w}
    },
\end{equation}
Eqn.~\eqref{eq:matrix_bc} gives
\begin{equation}
    S_{ZZ}^{\mathrm{BC}}(\mathbf{k})
    =
    \frac{
    m(\mathbf{k})
    }{
    1+\dfrac{\kappa_D^2}{k^2}m(\mathbf{k})
    }.
    \label{eq:szz_sm}
\end{equation}
Provided $m(\mathbf{k})$ has a finite nonzero limit as $k\rightarrow0$,
\begin{equation}
    S_{ZZ}^{\mathrm{BC}}(\mathbf{k})
    =
    \frac{k^2}{\kappa_D^2}
    +
    \mathcal{O}(k^4).
    \label{eq:szz_small_k}
\end{equation}
Thus the matrix form in Eqn.~\eqref{eq:matrix_bc} enforces the Stillinger--Lovett small-$k$ behavior in the charge channel,
while the individual elements of $\mathbf{S}^{\mathrm{BC}}(0^+)$ can stay finite. 

\subsection{Projection onto the charge-neutral composition manifold}

As detailed in Part I: Neutral Multi-component Mixtures~\cite{savoj2026computing},
the $\mathbf S^0$ matrix can be converted to the $\mathbf{U}$ matrix storing the partial derivative of chemical potentials with respect to particle numbers by
\begin{equation}
\mathbf{U} = \kb T\left[\mathbf{B}^{-1} -
\dfrac{\mathbf{B}^{-1}\mathbf{x}\mathbf{x}^T\mathbf{B}^{-1}}{\mathbf{x}^T\mathbf{B}^{-1}\mathbf{x}}\right],
\label{eq:muder}
\end{equation}
where $\mathbf{B}=\mathbf{Y} \mathbf S^0 \mathbf{Y}$.
In systems with charged components, the finite matrix $\mathbf{S}^{\mathrm{BC}}(0^+)$ is used in place of $\mathbf S^0$, and the pseudoinverse $\mathbf{B}^+$ over all nonzero eigenmodes is used instead of the direct inverse matrix $\mathbf{B}^{-1}$.

Similar to Part I, we then specify the projection used to express thermodynamic derivatives in a set of independent composition variables $\tilde{\mathbf{x}}$. 
From the vector of particle number fractions $\mathbf{x}$,
\begin{equation}
\mathbf{Q}\mathbf{x}
=
\begin{bmatrix}
\mathbf{P}_{f\times C} \\ \hline
\mathbf{1}_{1\times C} \\
\mathbf{z}_{1\times C}
\end{bmatrix}
\mathbf{x}
=
\begin{bmatrix}
\tilde{\mathbf{x}} \\ \hline
1 \\
0
\end{bmatrix},
\label{eq:selectionMatCharged}
\end{equation}
where $C$ is the number of components, $f = C-2$ is the number of independent particle number fractions, and $\mathbf{P}_{f\times C}$ is a projection matrix of rank $f$. $\mathbf{P}_{f\times C}$ can take many forms; the simplest choice has all diagonal elements $P_{ii}$ set to 1 and all other elements set to 0. The inverse mapping is 
\begin{equation}
    \mathbf{x}
    =
    \mathbf{Q}^{-1}
    \begin{bmatrix}
    \tilde{\mathbf{x}} \\ \hline
    1 \\
    0
    \end{bmatrix}.
    \label{eq:x_from_xtilde_charged}
\end{equation}
We then take the transformation matrix $\mathbf{M}$ to be the first $f$ columns of the inverse mapping,
\begin{equation}
    \mathbf{M}
    =
    \left(\mathbf{Q}^{-1}\right)_{[:,1:f]},
    \qquad
    \frac{\partial x_\alpha}{\partial \tilde{x}_\beta}
    =
    M_{\alpha \beta}.
    \label{eq:M_charged}
\end{equation}
By construction, $\mathbf{1}^{T}\mathbf{M}=\mathbf{0}^{T}$ and $\mathbf{z}^{T}\mathbf{M}=\mathbf{0}^{T}$, so infinitesimal changes $d\mathbf{x}=\mathbf{M}d\tilde{\mathbf{x}}$ remain on the normalized, charge-neutral composition manifold. 
The chemical potential derivatives with respect to the independent variables are
\begin{equation}
    \mathbf{\Gamma}
    =
    \mathbf{U}\mathbf{M},
    \qquad
    \Gamma_{\alpha\beta}
    =
    \frac{\partial\mu_{\alpha}}{\partial\tilde{x}_{\beta}},
    \label{eq:gamma_charged}
\end{equation}
which can be used to determine the chemical potential of a component $\alpha$ at a certain composition, $\mu_{\alpha}(\tilde{\mathbf{x}})$, by using Gaussian process integration over $\tilde{\mathbf{x}}$ as detailed in Part I.
The integration has better numerical behavior if performed on the excess chemical potentials
\begin{equation}
    \mu^{\mathrm{ex}}_{\alpha} = \mu_{\alpha} - k_B T\ln(x_\alpha),
    \label{eq:mu_ex}
\end{equation}
based on the partial derivatives
 \begin{equation}
      \Gamma^{\mathrm{ex}}_{\alpha \beta} 
      = \Gamma_{\alpha\beta} -k_BT \frac{1}{x_{\alpha}} M_{\alpha \beta}.
      \label{eq:gammaexeq}
 \end{equation}

\section{Aqueous electrolytes}

\begin{figure*}[t!]
    \centering
    \includegraphics[width=\textwidth]{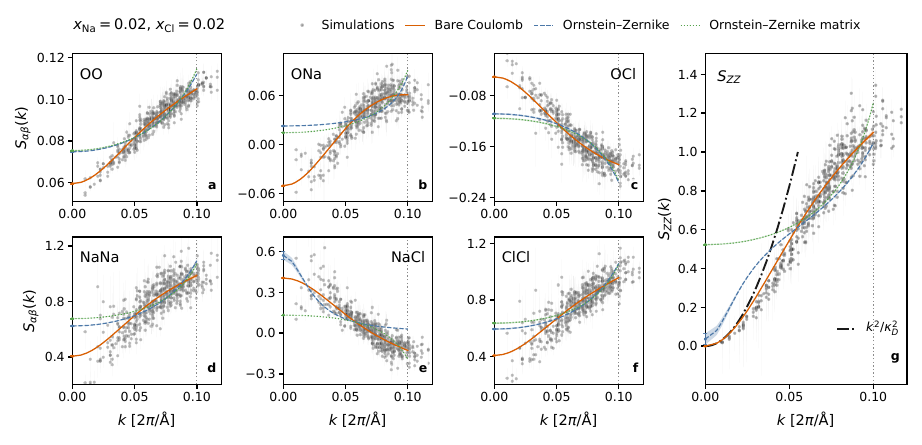}
    \caption{
Fitting partial structure factors for aqueous NaCl at $
x_{\mathrm{Na}} = x_{\mathrm{Cl}} = 0.02
$.
Panels (a--f) compare simulated partial structure factor components with the bare-Coulomb fit, the independent scalar Ornstein--Zernike fit, and the full Ornstein--Zernike matrix fit.
Panel (g) shows the corresponding charge--charge structure factor $S_{ZZ}(k)$, and the reference scaling curve $k^2/\kappa_D^2$. Statistical errors of the fits are indicated by the shaded bands, and the error bars at $k=0$.}

    \label{fig:naclwater-component-szz-fits}
\end{figure*}

We consider aqueous solutions containing NaCl and a second sodium halide, NaX, where X denotes either iodide (I$^-$) or fluoride (F$^-$). 
Anions with high charge density, such as fluoride, are generally classified as kosmotropes and tend to structure the surrounding water network, whereas larger and more polarizable anions, such as iodide, are more chaotropic and tend to disrupt it~\cite{Franz2004Zur,Zhao2016Protein}. 
Here we use the charge-aware S0 framework to compute salt chemical potentials in mixed-halide aqueous solutions.

We perform MD simulations of charge-neutral aqueous salt solutions at 298.15~K and 1~atm using LAMMPS~\cite{LAMMPS}. 
Each simulation contains around 32,000 water molecules, with Na$^+$ counterions added to neutralize the selected numbers of Cl$^-$ and X$^-$ ions, $x_{\mathrm{Na}}=x_{\mathrm{Cl}}+x_{\mathrm{X}}$. 
Each run lasts for 4~ns, using a 2~fs timestep, a Nos\'{e}--Hoover thermostat~\cite{Evans1985Nose} and barostat~\cite{shinoda2004rapid}. 
Interactions are described using the Joung--Cheatham ion parameters~\cite{joung2008determination} with rigid SPC/E water~\cite{berendsen1987missing} enforced by the SHAKE algorithm~\cite{ryckaert1977numerical}.

\subsection{NaCl in water solution}
\label{sec:NaClinWater}
We first perform a separate set of simulations and analyses for the NaCl in water edge of this composition space, 
in order to test the charge-aware small-$k$ extrapolation and chemical potential integration.
The chemical potentials for the NaCl-water system were calculated in Ref.~\cite{cheng2022computing}, using the neutral S0 method by combining Na$^+$ and Cl$^-$ ions as a single, uncharged component.

From each NPT trajectory, we compute the partial structure-factor matrix $\mathbf{S}(\mathbf{k})$ for the Na$^+$, Cl$^-$, and water oxygen at the reciprocal-cell wavevectors accessible in the simulation box. 
We then compare independent scalar Ornstein--Zernike (OZ) fits based on Eqn.~\eqref{eq:single_oz}, the coupled Ornstein--Zernike matrix  (OZ matrix) fit in Eqn.~\eqref{eq:matrix_oz},
and the bare-Coulomb matrix form (BC matrix) in Eqn.~\eqref{eq:matrix_bc}.
\figref{fig:naclwater-component-szz-fits} compares these fits using $k\le 0.1 \times 2\pi$/\AA, at $x_{\mathrm{Na}}=x_{\mathrm{Cl}}=0.02$, corresponding to $c_{\mathrm{NaCl}}=1.12$~mol~L$^{-1}$.
The bare-Coulomb fit uses a fixed Debye length, $\lambda_D=0.289$~nm ($\kappa_D=0.346$~\AA$^{-1}$), computed from the standard Debye expression of Eqn.~\eqref{eq:kappa_pure} using the experimental dielectric constant of water at room temperature.
The statistical errors of the fits are all quite small, with the independent OZ fit having the largest error bars amongst all. The small error bars are a result of the combined information from fits using multiple wavevectors and are computed using block averaging.

\figsrefsub{fig:naclwater-component-szz-fits}{a-f} show that the bare-Coulomb fit (orange curves) leads to good agreement with the simulated $\mathbf{S}(\mathbf{k})$ for all component pairs,
while the independent and Ornstein--Zernike matrix fits generally show large discrepancies at small-$k$ values.
More tellingly, \figref{fig:naclwater-component-szz-fits}(g) compares the three fits of the reconstructed charge-charge structure factor (Eqn.~\eqref{eq:szz_def}), illustrating that only the bare-Coulomb matrix fit can capture the suppression of macroscopic charge fluctuations and the Stillinger--Lovett limiting behavior~\cite{lee1997charge,stillinger1968ion}, $S_{ZZ}(k)\sim k^2 / \kappa_D^2$ at $k \rightarrow 0$.
Based on these observations, we thus conclude that
the bare-Coulomb matrix fit is the only form here appropriate for fitting the structure factors of charged components.

Next we convert the fitted $\mathbf{S}^0$ values to composition derivatives of excess chemical potentials by first using Eqn.~\eqref{eq:muder} to obtain $\mathbf{U}$,
and then converting $\mathbf{U}$ to the $\mathbf{\Gamma}$ matrix in Eqn.~\eqref{eq:gamma_charged} with the independent component $\widetilde{x} = x_{\mathrm{Na}}=x_{\mathrm{Cl}}$.
The excess derivatives are obtained using Eqn.~\eqref{eq:gammaexeq}, and integrated using GP to obtain the excess chemical potentials. 
Because individual ionic chemical potentials depend on the electrostatic convention 
and $\mu_{\mathrm{Na}}$ and $\mu_{\mathrm{Cl}}$ cannot be meaningfully disentangled from MD simulations with a fixed $x_{\mathrm{Na}}/x_{\mathrm{Cl}} = 1$ ratio,
we report the excess chemical potential (Eqn.~\eqref{eq:mu_ex}) of the neutral salt pair as $\mu_{\mathrm{NaCl}}^\mathrm{ex} = \mu^\mathrm{ex}_{\mathrm{Na}}+\mu^\mathrm{ex}_{\mathrm{Cl}}$.

\begin{figure}[t!]
    \centering
    \includegraphics[width=0.4\textwidth]{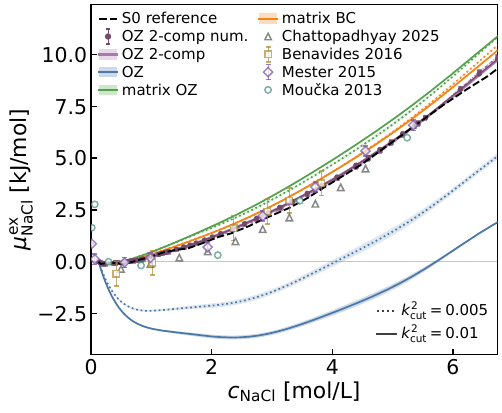}
\caption{
Excess chemical potential for NaCl salt in water,
$\mu_{\mathrm{NaCl}}^{\mathrm{ex}}(x)$, obtained by integrating chemical-potential derivatives inferred from fitted $\mathbf S^0$ values.
Dotted and solid curves illustrate the results obtained using different cutoff values of $k^2$ (units in $4\pi^2/\mathrm{\AA}^2$).
All curves are shifted to zero at the lowest simulated NaCl concentration.
Shaded bands indicate the estimated standard error.
Blue, green, and orange curves show results based on 
independent scalar Ornstein--Zernike fits, the coupled Ornstein--Zernike matrix fits,
and the bare-Coulomb matrix fits, respectively.
The purple circles and curves are based on treating $\mathrm{Na}^{+}$ and $\mathrm{Cl}^{-}$ ions as a single component and water as another. Literature results are included for comparison: the two-component S0 method (dashed black line)~\cite{cheng2022computing}, Kirkwood–Buff integration with finite-size corrections (gray triangles)~\cite{Chattopadhyay2025KB}, thermodynamic integration (yellow squares)~\cite{benavides2016consensus}, ion-pair insertion (purple diamonds)~\cite{Mester2015},  and osmotic-ensemble Monte Carlo (green circles)~\cite{mouvcka2013molecular}. }

    \label{fig:naclwater-mu-ex-validation}
\end{figure}

Fig.~\ref{fig:naclwater-mu-ex-validation} shows the
$\mu_{\mathrm{NaCl}}^\mathrm{ex}$ results obtained using the fitted $\mathbf S^0$ values based on the 
independent scalar Ornstein--Zernike, as well as the Ornstein--Zernike and the bare-Coulomb fits to the whole $\mathbf S(\mathbf{k})$ matrix.
The BC matrix and OZ matrix results show good agreement, even though the latter has far worse fits to the simulated $\mathbf S(\mathbf{k})$ values (Fig.~\ref{fig:naclwater-component-szz-fits}).
The reason is the cancellation of error, for example,
the OZ matrix fit overpredicts  $S_{\mathrm{ONa}}^0$  and underpredicts $S_{\mathrm{OCl}}^0$ by similar amounts.
In contrast, the individual OZ fits do not benefit from such cancellation and the resulting $\mu_{\mathrm{NaCl}}^\mathrm{ex}$ values show large deviations and are sensitive to the $k$-cutoff value used. 
Thus, agreement in an integrated thermodynamic quantity does not by itself establish that the underlying long-wavelength fluctuations have been represented correctly.

For comparison, 
the dashed black curve in Fig.~\ref{fig:naclwater-mu-ex-validation} shows the $\mu^{\rm ex}_{\rm NaCl}$ from Ref.~\cite{cheng2022computing},
which uses the original two-component S0 method.
It treats $\mathrm{Na}^{+}$ and $\mathrm{Cl}^{-}$ ions as a single salt component (S) and water as another component (O), and fits $S_\mathrm{OO}(\mathbf{k})$, $S_\mathrm{OS}(\mathbf{k})$, and $S_\mathrm{SS}(\mathbf{k})$ with the individual OZ fits.
The purple circles and the purple curves are based on the same protocol and use either direct numerical integration based on the trapezoidal rule or GP integration. 
All three approaches yield very similar results,
and are also consistent with the BC matrix-derived $\mu_{\mathrm{NaCl}}^\mathrm{ex}$.
Moreover, the S0-based results agree well with previous theoretical studies using Kirkwood--Buff integration with finite-size corrections~\cite{Chattopadhyay2025KB}, thermodynamic integration~\cite{benavides2016consensus}, ion-pair insertion~\cite{Mester2015}, and osmotic-ensemble Monte Carlo~\cite{mouvcka2013molecular}.
This confirms that treating the ion pair as a neutral component provides a valid route for this binary electrolyte.
However,
the extension of such an approach to more components is not straightforward; therefore it is more rigorous and general to apply the BC matrix fits.

\FloatBarrier

\subsection{Mixed-halide NaCl/NaX solutions}

Using a system size of 32,000 water molecules and ions and the same simulation procedure as for the NaCl-water solution above, we now determine how the chemical potentials of the salts change when adding a second halide, I$^-$ or F$^-$, with atomic fractions up to $x_\mathrm{I}=0.1$ and $x_\mathrm{F}=0.04$. We performed MD simulations using sizes of 4,000, 8,000, 16,000 and 32,000 for the total numbers of ions and water molecules and found that the $\mathbf{S^0}$ values showed weak system size dependence across this size range. We therefore concluded that a system size of 32,000 water molecules and ions is sufficient to converge the $\mathbf{S^0}$ values.
The BC fits use the same fixed-Debye-length protocol as in the NaCl-water validation, with $k_{\mathrm{cut}}=0.1 \times 2\pi/\mathrm{\AA}$.
The independent components are $\widetilde{\mathbf x} = [x_{\mathrm{Cl}},x_{\mathrm{X}}]^T$.
The gradients of excess chemical potentials with respect to $\widetilde{\mathbf{x}}$ are integrated using the Gaussian process procedure from Part I~\cite{savoj2026computing}, taking pure water as the reference point where the excess chemical potentials of all salts are set to zero.
Finally, we report neutral ion-pair salt combinations, e.g.,
$\mu^{\rm ex}_{\mathrm{NaX}}=\mu^{\rm ex}_{\mathrm{Na}}+\mu^{\rm ex}_{\mathrm{X}}$.

\begin{figure*}[t!]
    \centering
    \includegraphics[width=\textwidth]{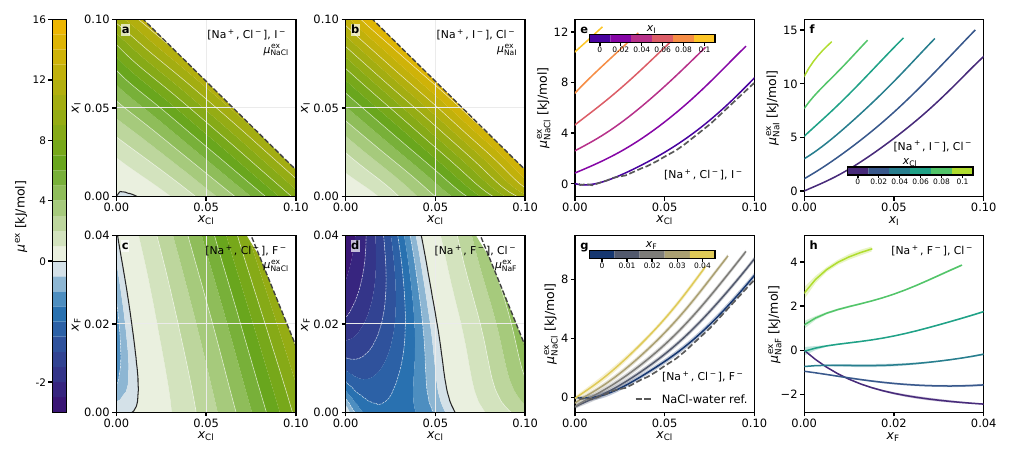}
\caption{
Excess chemical potentials for mixed aqueous salts, derived from the bare-Coulomb matrix fit.
Panels (a--d) show Gaussian process-integrated surfaces of the neutral ion-pair excess chemical potentials: $\mu^{\mathrm{ex}}_{\mathrm{NaCl}}$ and $\mu^{\mathrm{ex}}_{\mathrm{NaI}}$ for NaCl/NaI in water, and $\mu^{\mathrm{ex}}_{\mathrm{NaCl}}$ and $\mu^{\mathrm{ex}}_{\mathrm{NaF}}$ for NaCl/NaF in water.
Panels (e--h) show one-dimensional slices through the same surfaces; line colors indicate fixed $x_{\mathrm{I}}$ (e), fixed $x_{\mathrm{F}}$ (g), or fixed $x_{\mathrm{Cl}}$ (f, h), as marked by the inset color bars.
Shaded bands show standard error, and the dashed gray curve shows the reference values from NaCl-water simulations.
}
    \label{fig:bc-salt-mu-surfaces}
\end{figure*}

\figref{fig:bc-salt-mu-surfaces} summarizes the neutral ion-pair excess chemical potentials for both salt mixtures.
For more quantitative illustrations, 
\figsrefsub{fig:bc-salt-mu-surfaces}{e-h} show one-dimensional slices through these $\mu^{\mathrm{ex}}$ surfaces.
As validation, 
\figsrefsub{fig:bc-salt-mu-surfaces}{e,g} show that 
$\mu^{\mathrm{ex}}_{\mathrm{NaCl}}$ at $x_\mathrm{X}=0$ agrees well with reference values computed from the independent NaCl-water simulations from the previous section. Additionally, we find $\mu^{\rm ex}_{\rm NaI}(x)$ and  $\mu^{\rm ex}_{\rm NaF}(x)$ in NaCl-free solutions are consistent with previous ion-pair insertion~\cite{Mester2015Temperature} and individual-ion insertion~\cite{Saravi2021Individual} studies.

Salt excess chemical potentials are affected very differently by the addition of iodide and fluoride.
For NaCl/NaI in water, increasing either NaI at fixed $x_{\mathrm{Cl}}$ or NaCl at fixed $x_{\mathrm{I}}$ produces a steep increase in both $\mu^{\rm ex}_{\rm NaCl}$ and $\mu^{\rm ex}_{\rm NaI}$.
As such, \figrefsub{fig:bc-salt-mu-surfaces}{a} and \figrefsub{fig:bc-salt-mu-surfaces}{b} look similar and almost symmetric with respect to the $x_\mathrm{Cl}$ and $x_\mathrm{I}$ axes. In contrast, adding NaF over the sampled $x_{\mathrm{F}}\le 0.04$ decreases the excess chemical potentials of the NaCl and NaF salts at low NaCl concentrations, as shown by the negative blue region in \figsrefsub{fig:bc-salt-mu-surfaces}{c,d}, for both $\mu_{\mathrm{NaCl}}^{\mathrm{ex}}$ and $\mu_{\mathrm{NaF}}^{\mathrm{ex}}$.

The contrast with the NaI case is particularly clear when comparing the one-dimensional slices of $\mu_{\mathrm{NaCl}}^{\mathrm{ex}}$ in \figrefsub{fig:bc-salt-mu-surfaces}{e} and \figrefsub{fig:bc-salt-mu-surfaces}{g}: the $\mu_{\mathrm{NaCl}}^{\mathrm{ex}}$ curves separate strongly with increasing $x_{\mathrm{I}}$, but remain more closely grouped with increasing $x_{\mathrm{F}}$, particularly at low $x_\mathrm{Cl}$. 

At matched $x_X$ and $x_{\mathrm{Cl}}$, the solutions containing NaI  or NaF have similar ionic strengths and the same Na$^+$ fraction, yet $\mu_{\mathrm{NaCl}}^{\mathrm{ex}}$ is consistently higher in the NaI-containing solutions, 
demonstrating an anion-specific effect on NaCl activity that cannot be attributed simply to ionic strength or sodium concentration. The contrasting behavior is consistent with the different hydration characteristics of F$^-$ and I$^-$, and illustrates how ion-specific molecular interactions are reflected in the chemical potentials recovered by the S0 framework.

\begin{figure}[t!]
    \centering
    \includegraphics[width=0.4\textwidth]{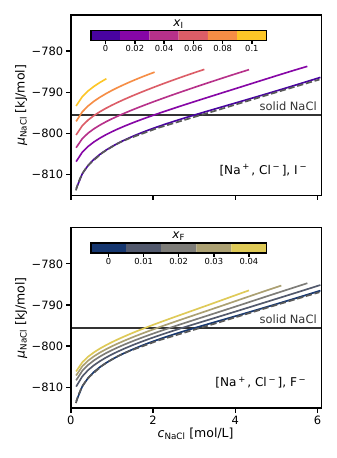}
\caption{
Chemical potential of NaCl in aqueous NaCl/NaI (top) and NaCl/NaF (bottom) mixtures, expressed per neutral ion pair. Colored curves show fixed co-anion mole fractions, obtained by GP integration of bare-Coulomb matrix $\mathbf{S^0}$ fits and addition of the ideal contribution. The gray dashed curve is the pure-NaCl/water S0 reference. The horizontal line marks the crystal chemical potential; intersections indicate solubility. Shading represents one standard error. 
}
    \label{fig:salt-solubility}
\end{figure}

To obtain the total NaCl chemical potential, we restore the ideal contributions of both ions to the GP-integrated excess chemical potentials, following Eq.~\eqref{eq:mu_ex}, and calibrate the additive constant using the reference solvation free energy from Ref.~\cite{reinhardt2023streamlined}. 
Relative to a pure-NaCl/water reference with $x_{\mathrm{Na}}=x_{\mathrm{Cl}}=x_0$, the ideal contribution is $k_\mathrm{B}T\ln(x_{\mathrm{Na}}x_{\mathrm{Cl}}/x_0^2)$, where charge neutrality requires $x_{\mathrm{Na}}=x_{\mathrm{Cl}}+x_X$. We use our previously determined solvation free energy of $-741.82\pm0.21$~kJ/mol
and crystal chemical potential of  $-795.565\pm0.063$~kJ/mol
at $298.15\,\mathrm{K}$ and 1~atm~\cite{reinhardt2023streamlined}, both expressed per NaCl ion pair. 
Fig.~\ref{fig:salt-solubility} compares the resulting solution chemical potentials at fixed iodide or fluoride mole fractions with the crystal value; their intersections determine NaCl saturation. 
The pure-water NaCl solubility is predicted to be approximately $3.0\,\mathrm{mol/L}$. 
For comparison, at $x_{\mathrm{I}}=0.04$ and $x_{\mathrm{F}}=0.04$, the estimated NaCl solubilities are $1.2$ and $1.8\,\mathrm{mol/L}$, respectively. 
A part of the NaCl solubility suppression simply comes from the additional Na$^+$ introduced when adding the second salt.
Nevertheless, the comparison shows that I$^-$ suppresses NaCl solubility more strongly than F$^-$ at the same co-anion mole fraction.

\section{Molten salts}
\label{sec:molten_salts}

Molten salts are important in clean-energy technologies, but their thermophysical properties are not characterized comprehensively because of the strong ionicity and the large composition space~\cite{roper2022molten}. 
The SuperSalt work~\cite{supersalt2025} introduced a data set~\cite{supersaltZenodo2025} covering the chemical space of
multi-component chloride molten salts,
and fitted a machine learning interatomic potential (MLIP) based on the MACE~\cite{batatia2022mace} architecture.

However, the SuperSalt MACE only contains short-range (SR) interactions, without explicitly considering electrostatics.
To include long-range electrostatics, we trained a MACELES MLIP on the same SuperSalt data set~\cite{supersaltZenodo2025} by combining the same short-range MACE setting~\cite{supersalt2025,supersaltZenodo2025} with the Latent Ewald Summation (LES) method~\cite{Cheng2025Latent,King2025Machine,zhong2025machine,Kim2025Universalb}, 
which infers partial charges and resulting long-range electrostatics just from energy and force training labels. 

We then evaluated both our MACELES model and the SuperSalt MACE-SR on the same training and testing data splits~\cite{supersalt2025},
and reported
the resulting force and energy root-mean-square errors (RMSEs) in Table~\ref{tab:supersalt-errors}.
The MACELES model gives lower force RMSEs and lower or comparable energy RMSEs than the published SuperSalt MACE-SR model for the training set, validation set, and two test sets each containing ternary and multi-component configurations. 

\begin{table*}
\caption{
Comparison of force and energy root-mean-square errors (RMSEs) for the MACELES model trained in this work and the published SuperSalt MACE-SR model~\cite{supersalt2025} evaluated on the SuperSalt data splits~\cite{supersaltZenodo2025}.
Force RMSEs are reported in meV/\AA, and energy RMSEs are reported in meV/atom.
\label{tab:supersalt-errors}
}
\begin{ruledtabular}
\begin{tabular}{lcccc}
Dataset &
\multicolumn{2}{c}{MACELES} &
\multicolumn{2}{c}{MACE-SR} \\
&
$F$ RMSE &
$E$ RMSE &
$F$ RMSE &
$E$ RMSE \\
\hline
Test 1 (multi-component) & 20.298 & 0.883 & 24.396 & 1.304 \\
Test 2 (ternary) & 13.606 & 0.591 & 17.590 & 0.585 \\
Training & 7.573 & 1.179 & 13.676 & 1.161 \\
Validation & 11.605 & 1.523 & 16.238 & 1.538 \\
\end{tabular}
\end{ruledtabular}
\end{table*}

All the NPT MD simulations of molten salts were performed at 1200~K and 1~bar using the Atomic Simulation Environment (ASE)~\cite{ase2017} with a system size of 6,400 atoms. Temperature and pressure were controlled using a Nos\'{e}–Hoover thermostat~\cite{Evans1985Nose} and a Martyna–Tobias–Klein barostat~\cite{martyna1994constant}, respectively. 
A 2~fs timestep was used with configurations saved every 0.4~ps after a 50~ps equilibration. The pure molten NaCl system and molten salt mixture systems had production run times of 500 ps and 200 ps, respectively. We tested time convergence by comparing radial distribution functions (RDFs) and fitted S0 from 100, 200, and 500~ps production runs. Since both values only weakly depend on simulation duration, we used 200-ps production runs for the molten salt mixtures.

\subsection{Molten NaCl}
\label{sec:molten_NaCl}

\begin{figure}[t!]
    \centering
    \includegraphics[width=0.4\textwidth]{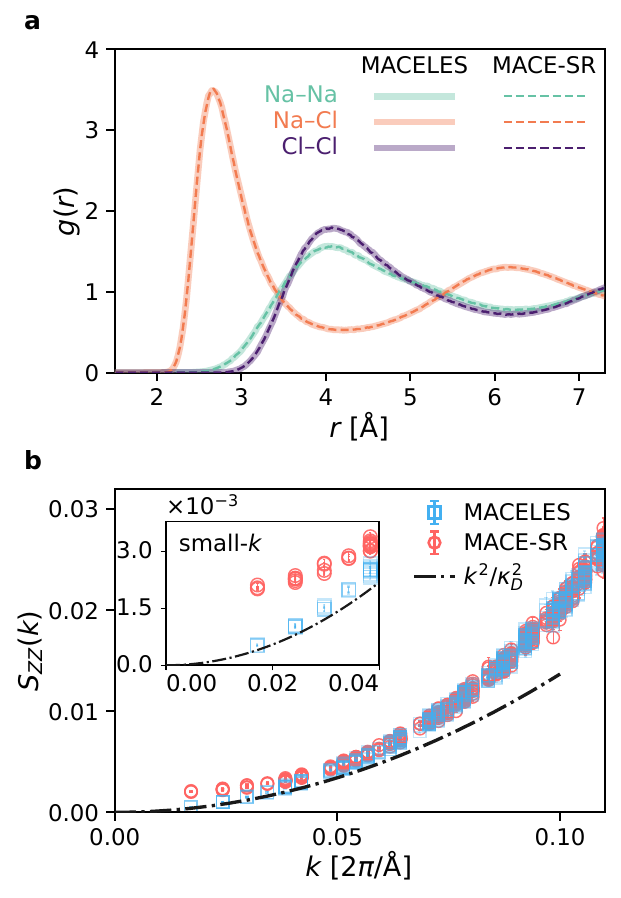}
\caption{Local and long-range structure of molten NaCl at 1200~K and 1~bar.
    \figLabelCapt{a} Radial distribution functions, $g(r)$, for Na--Na, Na--Cl and Cl--Cl, computed using MACELES (solid lines) and MACE-SR (dotted lines) models. Error bars denote standard errors from a time-block analysis.
    \figLabelCapt{b} Corresponding charge--charge structure factors, 
    $S_{\mathrm{ZZ}}(k)$, for MACELES (blue squares) and MACE-SR (red circles). 
    The reference curve $k^{2}/\kappa_{\mathrm{D}}^{2}$ is drawn with the Debye wavevector $\kappa_{\mathrm{D}}=5.37~\mathrm{\AA}^{-1}$. }
    \label{fig:benchamrk_NaCl}
\end{figure}

To assess the effect of long-range electrostatics, we compared MACELES with the original SuperSalt MACE-SR~\cite{supersalt2025}
using molten NaCl. 
\figrefsub{fig:benchamrk_NaCl}{a} shows that the radial distribution functions (RDFs) obtained with the two MLIPs are nearly indistinguishable, indicating that both predict very similar local arrangements.

\figrefsub{fig:benchamrk_NaCl}{b} compares the charge–charge structure factors, $S_{\rm ZZ}(\mathbf{k})$, from both models.
Specifically, $S_{\rm ZZ}(\mathbf{k})$ was computed using Eqn.~\eqref{eq:szz_def} with $z_{\rm Na}=+1$ and $z_{\rm Cl}=-1$. 
$\kappa_D$ was fixed at $5.37~\mathrm{\AA}^{-1}$, evaluated from Eqn.~\eqref{eq:matrix_kappa} with the experimental relative permittivity of $\varepsilon_{\infty}=1.95$ for molten NaCl at 1200~K and ambient pressure~\cite{Kochedykov2017RefractiveIndices}.
The MACELES $S_{\rm ZZ}(\mathbf{k})$ approaches the Stillinger--Lovett limiting form, whereas the MACE-SR result exhibits deviations at small $k$ values. 
Thus, similar local structure does not necessarily imply a correct collective electrostatic response: although MACELES and MACE-SR produce nearly indistinguishable RDFs, only MACELES recovers the expected long-wavelength charge behavior. We therefore use MACELES for the subsequent simulations.

\begin{figure}[htbp!]
    \centering
    \includegraphics[width=0.49\textwidth]{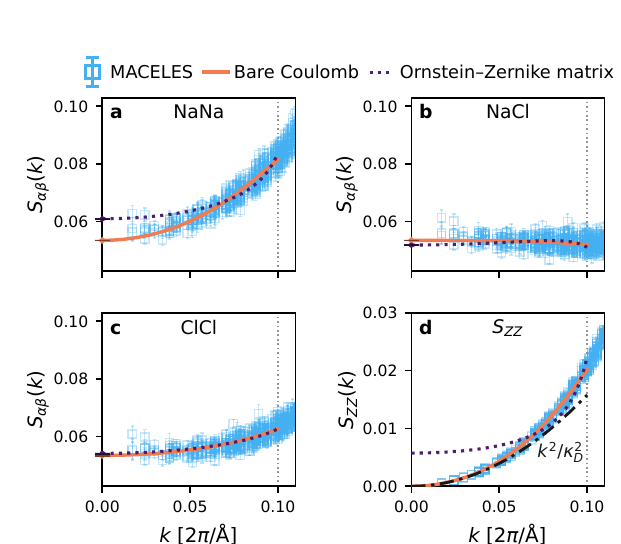}
    \caption{Partial structure factors for molten NaCl at 1200~K and 1~bar.
    \figLabelCapt{a--c:} MACELES MD results for the Na--Na, Na--Cl and Cl--Cl components, shown with the bare-Coulomb matrix fit (solid lines) and the full Ornstein--Zernike matrix fit (dotted lines). Fits were performed over $k^{2}<k_{\mathrm{cut}}^{2}=0.01\times 4\pi^2/\AA^2$, indicated by the vertical dotted lines. Error bars denote standard errors of simulated $S_{\alpha\beta}(k)$ from a block analysis. Diamonds (bare Coulomb) and circles (Ornstein--Zernike) at $k=0$ mark the corresponding $S_{\alpha\beta}(0)$, with error bars derived from 2000 Monte Carlo samples.
    \figLabelCapt{d:} Corresponding charge--charge structure factor $S_{\mathrm{ZZ}}(k)$. The dot-dashed line shows the Stillinger--Lovett limiting form $k^{2}/\kappa_{\mathrm{D}}^{2}$, using the fitted inverse Debye screening length $\kappa_{\mathrm{D}}=5.00~\mathrm{\AA}^{-1}$.
    }
    \label{fig:S0_NaCl}
\end{figure}

\figsrefsub{fig:S0_NaCl}{a--c} show the partial structure factors from the MACELES MD simulations, 
together with the bare-Coulomb fit (solid lines) and the Ornstein--Zernike matrix fit (dotted lines). 
Only the bare-Coulomb fit can accurately follow the small-$k$ MD results and follow the Stillinger--Lovett limiting behavior of $S_{\rm ZZ}(\mathbf{k})$.

To further validate the $S^0$ extrapolation using the bare-Coulomb fit,
we performed a thermodynamic consistency check: 
We computed the isothermal compressibility $\chi_T$,
\begin{equation}
    \chi_T  
    = -\dfrac{1}{V}
    \left( \dfrac{\partial V}{\partial P}\right)_{N,T},
    \label{eq:isothermal}
\end{equation}
using the relationship
\begin{equation}
    S^0_\mathrm{SS} = \rho_\mathrm{S} \kb T \chi_T,
    \label{eq:ssfrom_isothermal}
\end{equation}
where S denotes the Na$^+$ and Cl$^-$ ions treated as a single component, such that $S^0_\mathrm{SS}= \frac{1}{2}(S^0_\mathrm{NaNa}+S^0_\mathrm{ClCl}+2S^0_\mathrm{NaCl})$.
Using $k^2_{\rm cut}=0.01 \times 4\pi^2/\AA^2$ gives $S_{\rm SS}^0=0.106\pm0.004$ and $\chi_T=0.201\pm0.008$~GPa$^{-1}$. 
This is consistent with independent estimates of $\chi_T=0.219\pm0.006$~GPa$^{-1}$ from the fluctuation--dissipation relation~\cite{frenkel2023understanding} and $\chi_T=0.210\pm0.002$~GPa$^{-1}$ from a finite-difference evaluation of Eqn.~\eqref{eq:isothermal}. 
The close agreement among all three estimates validates the bare-Coulomb extrapolation and the resulting compressibility.

\subsection{Multi-component molten salt}

We applied the Coulomb-aware S0 method to compute the mixing free energies for LiCl--NaCl and \ce{MgCl2}--NaCl at 1200~K and 1~bar.  
For each mixture, we sampled 19 compositions, with LiCl component fraction $x_{\rm LiCl}=N_{\rm Li}/(N_{\rm Li}+N_{\rm Na})$ range from 0.05 to 0.95, and \ce{MgCl2} component fraction $x_{\rm MgCl_2}=N_{\rm Mg}/(N_{\rm Mg}+N_{\rm Na})$ range from 0.02 to 0.985.
Each composition was simulated using the same MD setup as molten NaCl described in Sec.~\ref{sec:molten_salts}. 
The partial structure-factor matrix, $\mathbf{S}(\mathbf{k})$, was then computed from 500 configurations and the $S^0$ values were obtained from the BC matrix fits in Eqn.~\eqref{eq:matrix_bc} with $k^2_{\rm cut}=0.02 \times 4\pi^2/\AA^2$. 
Since charge neutrality and particle fraction normalization leave a single degree of freedom in these three-element melts, we took $\tilde{x}=x_{\rm Li}$ and $\tilde{x}=x_{\rm Mg}$ as the independent fractions for LiCl--NaCl and \ce{MgCl2}--NaCl, respectively. In the subsequent GP integration, both systems used a stationary RBF kernel with a length scale of $\theta=0.1$, gradient noise level of $\sigma_g = 10^{-3}$~kJ/mol, and zero function noise level of $\sigma_f = 0$~kJ/mol.
The GP-integrated ionic chemical potentials were converted to neutral-salt combinations, i.e., $\mu_{\rm LiCl}=\mu_{\rm Li}+\mu_{\rm Cl}$, $\mu_{\rm NaCl}=\mu_{\rm Na}+\mu_{\rm Cl}$, and $\mu_{\rm MgCl_2}=\mu_{\rm Mg}+2\mu_{\rm Cl}$. 
The free energy of mixing $\Delta G^{\rm mix}$ for the two mixtures was computed as $x_{\rm LiCl}\mu_{\rm LiCl} + x_{\rm NaCl}\mu_{\rm NaCl}$, or $x_{\rm MgCl_2}\mu_{\rm MgCl_2} + x_{\rm NaCl}\mu_{\rm NaCl}$, respectively.

\begin{figure}[htbp!]
    \centering
    \includegraphics[width=0.4\textwidth]{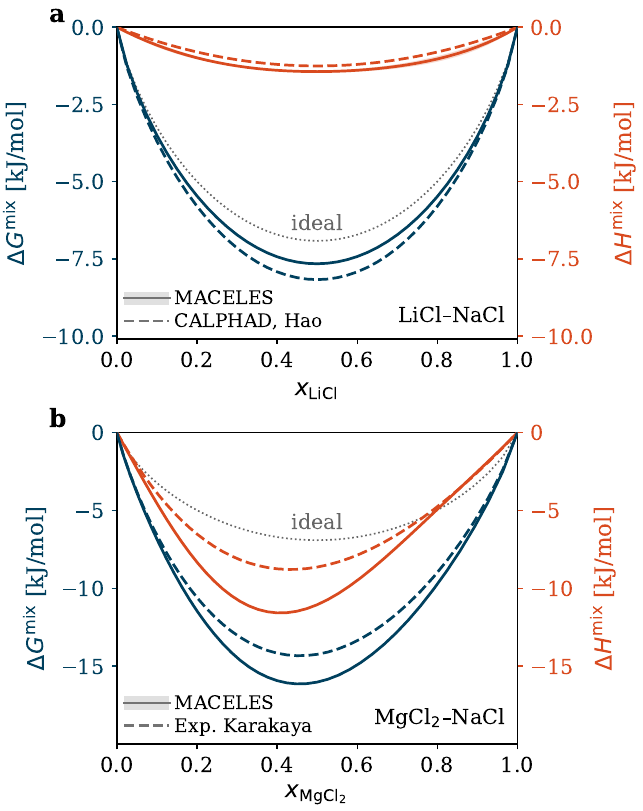}
    \caption{Free energy of mixing, $\Delta G^{\rm mix}$, and enthalpy of mixing
    $\Delta H^{\rm mix}$ for molten chloride mixtures at 1200~K and 1~bar.  
    \figLabelCapt{a}: $\Delta G^{\rm mix}$ (blue solid lines) and $\Delta H^{\rm mix}$ (orange solid lines) for LiCl--NaCl as a function of $x_{\rm LiCl}=N_{\rm Li}/(N_{\rm Li}+N_{\rm Na})$, compared with the previous CALPHAD model based on the regular-solution assumption (dashed lines)~\cite{Hao2023Thermodynamic}. 
    \figLabelCapt{b}: $\Delta G^{\rm mix}$ and $\Delta H^{\rm mix}$ for \ce{MgCl2}--NaCl as a function of $x_{\rm MgCl_2}=N_{\rm Mg}/(N_{\rm Mg}+N_{\rm Na})$, compared with previous electromotive-force measurements (dashed lines)~\cite{Karakaya1986Thermodynamic}. The line widths of the $\Delta G^{\rm mix}$ and $\Delta H^{\rm mix}$ curves indicate the statistical error propagated from the $\mathbf{S}(\mathbf{k})$ matrix and from the calculated enthalpy, respectively.}
    \label{fig:Gmix}
\end{figure}

The resulting $\Delta G^{\rm mix}$ (solid blue lines) shown in \figref{fig:Gmix} demonstrate distinct thermodynamic behaviors in the two mixtures. 
In contrast to LiCl--NaCl, NaCl--\ce{MgCl2} deviates strongly from ideal mixing (dotted gray line), with $\Delta G^{\rm mix}$ reaching $-16.1$~kJ/mol at $x_{\rm MgCl_2}=0.45$ against $-6.9$~kJ/mol for an ideal solution.  
This difference is further evidenced in the mixing enthalpies, $\Delta H^{\rm mix}$ (solid orange lines), calculated directly from the same NPT trajectories.
In particular, the minimum excess mixing enthalpy is approximately $-11.6$~kJ/mol for \ce{MgCl2}--NaCl, but only $-1.4$~kJ/mol for LiCl--NaCl.
These results indicate that the enhanced thermodynamic stability of the \ce{MgCl2}--NaCl mixture relative to ideal mixing is predominantly enthalpy-driven. 
Moreover, both simulated $\Delta G^{\rm mix}$ and $\Delta H^{\rm mix}$ agree well with the CALPHAD model~\cite{Hao2023Thermodynamic} for LiCl--NaCl and with previous electromotive force measurements~\cite{Karakaya1986Thermodynamic}
for \ce{MgCl2}--NaCl. 
The remaining discrepancies between our simulation results and experimental data may partly reflect limitations of the PBE-D3 level of theory used to generate the SuperSalt dataset~\cite{supersalt2025}, as previous density functional theory studies have reported that the thermophysical properties of molten salts are sensitive to the choice of exchange--correlation functional and dispersion correction~\cite{Corradini2014,Nam2014,attarian2025best}.

\begin{figure}[htbp!]
    \centering
    \includegraphics[width=0.49\textwidth]{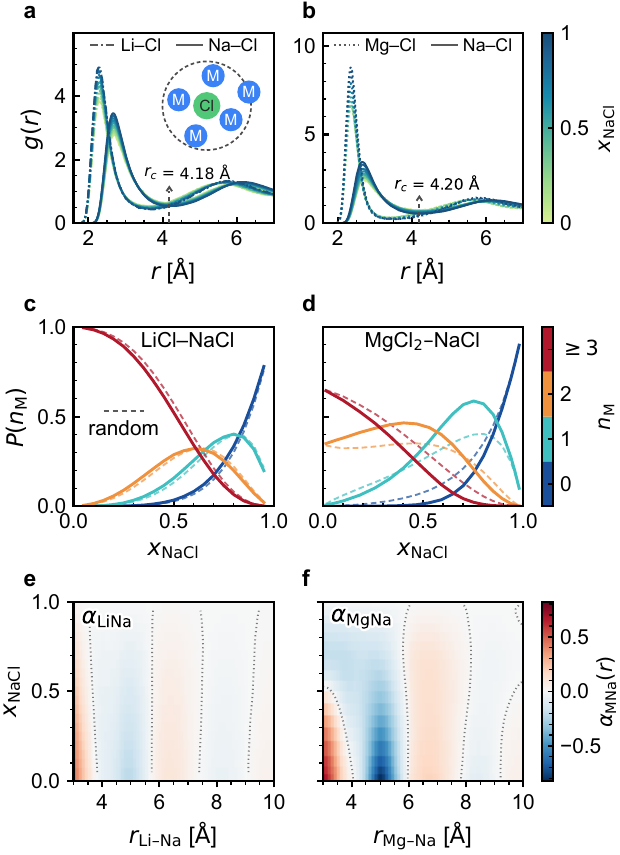}
    \caption{Composition-dependent local structure and spatial ordering in molten chloride salts at 1200~K and 1~bar.
    \figLabelCapt{a, b}: Cl-centered partial radial distribution functions, $g(r)$, for LiCl--NaCl (a) and \ce{MgCl2}--NaCl (b). The curves are colored by the NaCl fraction 
    $x_{\rm NaCl}=N_{\rm Na}/(N_{\rm Na}+N_{\rm M})$, M = Li or Mg. 
    The vertical gray lines indicate the first-coordination-shell cutoffs.
    \figLabelCapt{c, d}: Probability $P(n_{\rm M})$ that a Cl ion has $n_{\rm M}=0$, 1, 2, or $\geq3$ neighboring M cations within its first coordination shell, where M=Li (c) or Mg (d). 
    Solid lines show the simulated distributions, colored by $n_{\rm M}$, whereas dashed lines show the corresponding hypergeometric distributions expected for statistically random cation arrangements.
    \figLabelCapt{e, f}: Warren--Cowley parameters $\alpha_{\rm LiNa}(r)$ for LiCl--NaCl (e) and $\alpha_{\rm MgNa}(r)$ for ~\ce{MgCl2}--NaCl (f), shown as functions of the cation--cation separation $r$ and $x_{\rm NaCl}$. Positive values (red) indicate Na-depleted M-centered coordination shells, whereas negative values (blue) indicate Na-enriched shells. Dotted gray contours correspond to no preference.}
    \label{fig:connectivity}
\end{figure}

To probe the microscopic origin of the contrasting mixing behaviors, \figref{fig:connectivity} compares local cation ordering in the two melts.
~\figsrefsub{fig:connectivity}{a,b} show the Cl-centered cation--anion RDFs, which are later used to define the first coordination shell of Cl, with cutoff radii of $r_c=4.18$~\AA{} for LiCl--NaCl and $4.20$~\AA{} for \ce{MgCl2}--NaCl. 
We then counted the number of M cations ($\rm M=\mathrm{Li}$ or $\mathrm{Mg}$) within the first coordination shell of each Cl to quantify deviations of the local cation environment from random mixing.

\figsrefsub{fig:connectivity}{c,d} show the probability distribution $P(n_{\rm M})$ for the two melts, where $n_{\rm M}$ is the number of neighboring cations of type $\mathrm{M}$ ($\rm M=\mathrm{Li}$ or $\mathrm{Mg}$) in the coordination shell of an anion.
To quantify competition for anion coordination between the two cations ($\mathrm{M}$ and $\mathrm{M}^{\prime}$), we construct a random-mixing reference distribution ($P_{\rm rand}(n_{\mathrm M}|l)$, dashed line) by randomly permuting the cation identities while keeping the coordination fixed.
For an anion coordinated by $l$ cations, the conditional probability of finding $n_{\mathrm M}$ of type M and thus $l - n_{\mathrm M}$ of type $\mathrm{M}^{\prime}$ under random mixing is hypergeometric:
\begin{equation}
P_{\rm rand}(n_{\mathrm M}|l) = \frac{\binom{N_{\mathrm M}}{n_{\mathrm M}} \binom{N_{\mathrm M^{\prime}}}{l-n_{\mathrm M}} }{ \binom{N_{\mathrm M}+N_{\mathrm M^{\prime}}}{l}},
\end{equation}
where $N_{\mathrm M}$ and $N_{\mathrm M^{\prime}}$ are the total numbers of cations of types $\mathrm{M}$ and $\mathrm{M}^{\prime}$, respectively, and $\binom{\cdot}{\cdot}$ denotes the binomial coefficient.

\figrefsub{fig:connectivity}{c} shows that LiCl--NaCl closely follows the random-mixing reference across the full composition range.
In contrast, \figrefsub{fig:connectivity}{d} reveals systematic deviations from random mixing in \ce{MgCl2}--NaCl: Cl ions are typically depleted in environments with three or more Mg neighbors ($P(n_{\rm Mg}\geq3)<P_{\rm rand}(n_{\rm Mg}\geq3)$), and enriched in those with one or two Mg neighbors ($P(0<n_{\rm Mg}<3)>P_{\rm rand}(0<n_{\rm Mg}<3)$).
Consequently, adding NaCl fragments the extended Mg--Cl network while intensifying the first Mg--Cl RDF peak (see~\figrefsub{fig:connectivity}{b}), indicating a shorter mean Mg--Cl distance than that in pure \ce{MgCl2}. 
The shorter cation--anion separation strengthens their Coulombic interactions, which may contribute to the strongly exothermic $\Delta H^{\rm mix}$.
The inferred reorganization into less-connected Mg-centred units in \ce{MgCl2}--NaCl is consistent with the disruption of Mg-templated networks observed in combined X-ray scattering and MD studies of KCl-diluted molten \ce{MgCl2}~\cite{wu2019elucidating,wu2020temperature,roy2021unraveling}.

To complement the picture of NaCl-induced structural reorganization revealed by cation--anion coordination, we next examine the corresponding changes in cation--cation ordering for the two molten salts.  
\figsrefsub{fig:connectivity}{e,f} quantify the spatial ordering between cations as a function of the cation--Na distance $r_{\rm M-Na}$ using the Warren--Cowley parameter~\cite{Cowley1950Approximate,warren1990x}:
\begin{equation}
\alpha_{\rm MNa}(r)=1-P({\rm Na}|{\rm M},r)\,[N_{\rm Na}/(N_{\rm Na} + N_{\rm M} -1)]^{-1},
\end{equation}
where $P({\rm Na}|{\rm M},r)$ is the conditional probability of finding Na cations at distance $r$ from an M cation.
Accordingly, positive (negative) $\alpha_{\rm MNa}(r)$ values mark Na-depleted (M-enriched) and Na-enriched (M-depleted) shells, respectively. 
In LiCl--NaCl, $\alpha_{\rm LiNa}(r)$ remains relatively close to zero across all compositions (\figrefsub{fig:connectivity}{e}), indicating a nearly random cation distribution that accounts for its near-ideal mixing behavior. 
By contrast, the amplitude of $\alpha_{\rm MgNa}(r)$ decreases as $x_{\rm NaCl}$ increases. 
 The fading positive (red) regions in~\figrefsub{fig:connectivity}{f} show that NaCl addition progressively weakens the Mg--Mg spatial correlations in the mixture, suggesting substantially weaker Mg--Mg repulsion, which is consistent with the reduced connectivity of Mg-centered units inferred from \figrefsub{fig:connectivity}{d}. 
This reduced repulsion may help stabilize the mixture and provides a microscopic basis for the strongly negative $\Delta H^{\rm mix}$ and $\Delta G^{\rm mix}$ of \ce{MgCl2}--NaCl.

\section{Conclusions}
We extended the multi-component S0 framework to calculate the chemical potentials of charged mixtures from equilibrium NPT simulations by combining a bare-Coulomb matrix extrapolation of the $k \rightarrow 0$ limit with a matrix projection that enforces composition closure and charge neutrality. 
We benchmarked the framework using mixed aqueous halide salts, and molten chlorides containing both monovalent and divalent ions.

In contrast to the Ornstein--Zernike form, which cannot capture small-$k$ behavior of charged systems, the bare-Coulomb matrix form in Eqn.~\eqref{eq:matrix_bc} correctly describes the simulated partial structure factors and recovers the Stillinger--Lovett limit (Figs.~\ref{fig:naclwater-component-szz-fits} and~\ref{fig:S0_NaCl}). 
For molten NaCl, we also find that accurately describing collective charge fluctuations and thus the Stillinger--Lovett limiting behavior requires potentials with explicit long-range electrostatics. 
Moreover, the ~\ce{MgCl2}--NaCl benchmark demonstrates that the extended S0 framework applies directly to multivalent ionic systems without further modifications.

Compared with direct real-space KB integration, the extended S0 method uses a physically constrained bare-Coulomb matrix fit to extrapolate simulated partial structure factors to $k\rightarrow0$. This treatment avoids the separate RDF and integration-volume corrections used in real-space KB calculations~\cite{CortesHuerto2016,kruger2018size,Dawass2018KB,Dawass2019KB}. Moreover, the good agreement between the S0-derived excess chemical potential of aqueous NaCl and recent KB-based results~\cite{Chattopadhyay2025KB}, shown in Fig.~\ref{fig:naclwater-mu-ex-validation}, further validates the S0 framework.

Besides validating the S0 method, the two systems illustrate complementary
aspects of charged-mixture thermodynamics. In aqueous mixed-halide solutions, the
calculated chemical potentials distinguish the effects of F$^-$ and I$^-$ even at comparable ionic
strength and sodium concentration, demonstrating sensitivity to ion-specific solution environments. In molten NaCl, models with nearly identical local radial distribution
functions exhibit different long-wavelength charge fluctuations, showing that agreement in local structure alone does not guarantee the correct collective electrostatic response.

Although the present benchmarks contain at most two independent composition variables, the S0 framework can be extended to more complex mixtures. The uncertainty-guided CUR sampling introduced in Part I~\cite{savoj2026computing} provide a practical route for efficiently sampling the chemical–potential surface in higher-dimensional composition spaces.
The method developed here thus enables predictions of chemical potentials in complex ionic mixtures, supporting applications in solubility, ion exchange, and electrochemical materials design.

\textbf{Acknowledgments}
The authors acknowledge the research computing facilities provided by BRC UCB and LRC Lawrence Berkeley National Laboratory. 

\textbf{Data availability statement}
Empirical force-field parameters, fitted MLIP, MD input scripts, MD results, and data-analysis scripts generated for this study are available in the public repository: \url{https://github.com/ChengUCB/extended_S0}.

\textbf{Code availability} 
The Gaussian process regression code is publicly available at \url{https://github.com/ChengUCB/GPR_grad}. The multi-component S0 code is publicly available at \url{https://github.com/ChengUCB/S0_multi/}.

\bibliographystyle{apsrev4-1}

\end{document}